\documentstyle[twoside,fleqn,espcrc2,rotate,epsf]{article}

\newcommand{\AmS}{{\protect\the\textfont2
  A\kern-.1667em\lower.5ex\hbox{M}\kern-.125emS}}

\def\Preprint{\vspace*{-7.5cm} \noindent hep-ph/9709441 \\
 \noindent FTUV/97-44 \\ 
  IFIC/97-45 \\  
  \vspace{4.9cm}}   
\def\refjl#1#2#3#4#5#6{\bibitem{#1} #2, {#3} {#4} (#5) #6.}

\def\etal{{et al}}
\def\NPPS{Nucl. Phys. B (Proc. Suppl.)}
\def\PL{Phys. Lett.}
\def\PRL{Phys. Rev. Lett.}
\def\PR{Phys. Rev.}
\def\PRep{Phys. Rep.}
\def\ZP{Z. Phys.}

\def\RPP{Rep. Prog. Phys.}
\def\ARNPS{Ann. Rev. Nucl. Part. Sci.}
\def\PTP{Progr. Theor. Phys.}

\newcommand{\eqn}[1]{(\ref{#1})}
\newcommand{\be}{\begin{equation}}
\newcommand{\ee}{\end{equation}}
\newcommand{\no}{\nonumber}
\newcommand{\bel}[1]{\be\label{#1}}
\newcommand{\ba}{\begin{array}{c}}
\newcommand{\bat}{\begin{array}{cc}}
\newcommand{\ea}{\end{array}}
\newcommand{\beqn}{\begin{eqnarray}}
\newcommand{\eeqn}{\end{eqnarray}}

\newcommand{\bi}{\begin{itemize}}
\newcommand{\ei}{\end{itemize}}

\newcommand{\bV}{\mbox{\boldmath $V$}}
\newcommand{\rms}{\rm\scriptsize}

\newcommand{\e}{\mbox{\rm e}}
\newcommand{\bnul}{\stackrel{{}_{(-)}}{\nu_l}}

\newcommand{\cL}{{\cal L}}

\newcommand{\cM}{{\cal M}}
\newcommand{\cO}{{\cal O}}

\newcommand{\cH}{{\cal H}}

\newcommand{\cI}{{\cal I}}

\title{Weak Decays, Quark Mixing and CP Violation: Theory Overview
\thanks{
    Invited talk at the XVI International Workshop
    on Weak Interactions and Neutrinos (WIN'97), Capri,   
    June 1997}}

\author{A. Pich    \\ \noindent  
         Departament de F\'{\i}sica Te\`orica, 
         IFIC,  CSIC --- Universitat de Val\`encia, \\ 
         Dr. Moliner 50, E--46100 Burjassot, Val\`encia, Spain}
       
\begin{document}

\begin{abstract}
A brief overview of flavour--changing phenomena is presented.
The main topics discussed are
the universality and Lorentz structure
of the leptonic charged--current couplings,
our present knowledge of the quark--mixing matrix 
and the future prospects for CP--violation studies.
\end{abstract}


\maketitle
\Preprint
\section{INTRODUCTION}
\label{sec:introduction}

In spite of its enormous phenomenological success, the 
Standard Model (SM) leaves too many
unanswered questions to be considered as a complete description of the
fundamental forces.
We do not understand yet why fermions are replicated in three
(and only three)
nearly identical copies? Why the pattern of masses and mixings
is what it is?  Are the masses the only difference among the three
families? What is the origin of the SM flavour structure?
Which dynamics is responsible for the observed CP violation?

The fermionic flavour is the main source of
arbitrary free parameters in the SM: 9 fermion masses,
3 mixing angles and 1 complex phase (assuming the neutrinos to be
massless).
The problem of fermion--mass
generation is deeply related with the mechanism responsible for the 
electroweak spontaneous symmetry breaking.
Thus, the origin of these parameters lies in the most obscure part of
the SM Lagrangian: the scalar sector. 
Clearly, the dynamics of flavour appears to be ``terra incognita''
which deserves a careful investigation.

The flavour structure looks richer in the quark sector,
where mixing phenomena among the different families occurs
(leptons would also mix if neutrino masses were non-vanishing).
A precise measurement of the quark mixings would allow to test
their predicted unitarity structure, and could give some hints
about the unknown underlying dynamics.
Since quarks are confined within hadrons,
an accurate determination of their mixing parameters requires
first a good understanding of hadronization effects
in flavour--changing transitions.
The interplay of strong interactions in weak decays
plays a crucial role, which,
unfortunately, is   
difficult to control due to the
non-perturbative character of QCD at long distances.

In the SM flavour--changing transitions occur only in the charged--current
sector:
\beqn\label{eq:cc_mixing}
\lefteqn{\cL_{\mbox{\rms CC}}  =  {g\over 2\sqrt{2}}\,\left\{
W^\dagger_\mu\, J^\mu_W + h.c. \right\} \, ,}
\\
\lefteqn{J^\mu_W  = 
\sum_{ij}\,
\bar u_i\gamma^\mu(1-\gamma_5) \bV_{\!\! ij} d_j 
 +\sum_l\, \bar\nu_l\gamma^\mu(1-\gamma_5) l
\, .}
\no\eeqn
The so-called Cabibbo--Kobayashi--Maskawa \cite{CA:63,KM:73} (CKM)
matrix {\boldmath $V$} couples any {\it up--type} quark with all
{\it down--type} quarks.
It originates from the same (unknown) Yukawa couplings giving rise to the
quark masses.

For $N_G$ fermion generations, the quark--mixing matrix
contains $(N_G-1)^2$ physical parameters:
$N_G(N_G-1)/2$ moduli and $(N_G-1)(N_G-2)/2$ phases.
In the simpler case of two generations, {\boldmath $V$}
is determined by a single parameter, the so-called
Cabibbo angle \cite{CA:63}.
With $N_G=3$, the CKM matrix is described by 3 angles and 1 phase
\cite{KM:73}.
This CKM phase is
the only complex phase in the SM
Lagrangian; thus, it is a unique source of CP violation.
In fact, it was for this reason that the third generation
was assumed to exist \cite{KM:73}, 
before the discovery of the $\tau$ and the $b$ .
With two generations, the SM could not explain the observed
CP violation in the $K$ system.

\section{LEPTONIC DECAYS}

\begin{figure}[bth]
\epsfxsize =4.5cm \epsfbox{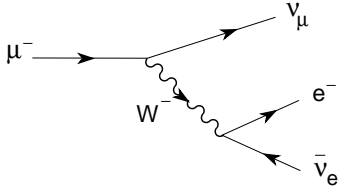}
\vspace{-0.5cm}
\caption{$\mu$--decay diagram. \label{fig:mu_decay}}
\end{figure}

The simplest flavour--changing process is the leptonic
decay of the muon, which proceeds through the $W$--exchange
diagram shown in Fig.~\ref{fig:mu_decay}.
The momentum transfer carried by the intermediate $W$ is very small
compared to $M_W$. Therefore, the vector--boson propagator reduces
to a contact interaction.
The decay can then be described through an effective local
4--fermion Hamiltonian,
\bel{eq:mu_v_a}
\cH_{\mbox{\rms eff}}\, = \, {G_F \over\sqrt{2}}
\left[\bar e\gamma^\alpha (1-\gamma_5) \nu_e\right]\,
\left[ \bar\nu_\mu\gamma_\alpha (1-\gamma_5)\mu\right]\, , 
\ee
where
$G_F/\sqrt{2} = g^2/ (8 M_W^2)$
is called the Fermi coupling constant.
$G_F$ is fixed by the total $\mu$--decay width,
\bel{eq:mu_lifetime}
\Gamma(\mu^-\to e^-\bar\nu_e\nu_\mu)
 =  {G_F^2 m_\mu^5\over 192 \pi^3}\,
f\left(m_e^2/m_\mu^2\right) \, r_{\mbox{\rms EW}} \, ,
\ee
where
$\, f(x) = 1-8x+8x^3-x^4-12x^2\ln{x}$,
and
$r_{\mbox{\rms EW}} = 0.9958$
takes into account the leading higher--order corrections \cite{KS:59}.
The measured $\mu$ lifetime \cite{pdg:96},
implies the value
$G_F =  (1.16639\pm 0.00002)\times 10^{-5} \:\mbox{\rm GeV}^{-2}$

\subsection{Lepton Universality}

The decays of the $\tau$ lepton proceed through the same
$W$--exchange mechanism as the leptonic $\mu$ decay.
The only difference is that several final states
are kinematically allowed:
$\tau^-\to\nu_\tau e^-\bar\nu_e$,
$\tau^-\to\nu_\tau\mu^-\bar\nu_\mu$,
$\tau^-\to\nu_\tau d\bar u$ and $\tau^-\to\nu_\tau s\bar u$.
Owing to the universality of the $W$ couplings, all these
decay modes have equal probabilities (if final fermion masses and
QCD interactions are neglected), except for an additional
$N_C |\bV_{\!\! ui}|^2$ factor ($i=d,s$) in the semileptonic
channels, where $N_C=3$ is the number of quark colours.

\begin{figure}[bth]
\centerline{\epsfxsize =7.5cm \epsfbox{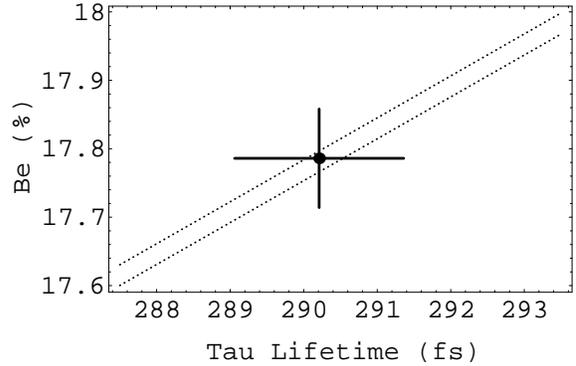}}
\vspace{-0.5cm}
\caption{$B_e$ / $\tau_\tau$ correlation
\protect\cite{tau:96}. The dotted
band is the SM prediction in Eq.~(\protect\ref{eq:relation}).
\label{fig:BeLife}}
\end{figure}

Using the value of $G_F$  
measured in $\mu$ decay, Eq.~\eqn{eq:mu_lifetime}
(with trivial kinematical changes)
provides a
relation  \cite{tau:96} between the $\tau$ lifetime
and the leptonic branching ratios
$B_l\equiv B(\tau^-\to\nu_\tau l^-\bar\nu_l)$:
\beqn
\label{eq:relation}
B_e & = & {B_\mu \over 0.972564\pm 0.000010} 
\no\\ & = &
{ \tau_{\tau} \over (1.6321 \pm 0.0014) \times 10^{-12}\, {\rm s} } \, .
\eeqn
%

The predicted $B_\mu/B_e$ ratio is in perfect agreement with the measured
value $B_\mu/B_e = 0.974 \pm 0.006$.  As shown in
Fig.~\ref{fig:BeLife}, the relation between $B_e$ and
$\tau_\tau$ is also well satisfied by the present data.
These measurements test the universality of
the $W$ couplings to the leptonic charged currents.
The $B_\mu/B_e$ ratio constraints $|g_\mu/g_e|$, while
$B_e/\tau_\tau$
provide information on $|g_\tau/g_\mu|$.
The present results  \cite{tau:96} are shown in Tables \ref{tab:univme} and
\ref{tab:univtm}, together with the values obtained from the
ratios
$R_{\pi\to e/\mu}\equiv\Gamma(\pi^-\to e^-\bar\nu_e)/
\Gamma(\pi^-\to\mu^-\bar\nu_\mu)$ 
and 
$R_{\tau/P}\equiv\Gamma(\tau^-\to\nu_\tau P^-)/
\Gamma(P^-\to \mu^-\bar\nu_\mu)$  
[$P=\pi,K$],
and from the comparison of the $\sigma\cdot B$ partial production
cross-sections for the various $W^-\to l^-\bar\nu_l$ decay
modes at the $p$-$\bar p$ colliders.

\begin{table}[bth]
\caption{Present constraints \protect\cite{tau:96} on $|g_\mu/g_e|$.}
\label{tab:univme}
\vspace{0.2cm}
\begin{tabular}{lc}
\hline
& $|g_\mu/g_e|$ \\ \hline
$B_\mu/B_e$ & $1.0005\pm 0.0030$
\\
$R_{\pi\to e/\mu}$ & $1.0017\pm 0.0015$
\\
$\sigma\cdot B_{W\to\mu/e}$ & $1.01\pm 0.04$
\\ \hline
\end{tabular}\vspace{1cm}
%
\caption{Present constraints \protect\cite{tau:96} on $|g_\tau/g_\mu|$.}
\label{tab:univtm}
\vspace{0.2cm}
\begin{tabular}{lc}
\hline
& $|g_\tau/g_\mu|$  \\ \hline
$B_e\tau_\mu/\tau_\tau$ & $1.0001\pm 0.0029$
\\
$R_{\tau/\pi}$ &  $1.005\pm 0.005$
\\
$R_{\tau/K}$ & $0.984\pm 0.020$
\\
$R_{\tau/h}$ & $1.004\pm 0.005$
\\
$\sigma\cdot B_{W\to\tau/\mu}$ & $0.99\pm 0.05$
\\ \hline
\end{tabular}
\end{table}
%


\subsection{Lorentz structure}

Let us consider the leptonic 
decay $l^-\to\nu_l l'^-\bar\nu_{l'}$. 
The most general, local, derivative--free, lepton--number conserving, 
four--lepton interaction Hamiltonian, 
consistent with locality and Lorentz invariance,
%
\be
{\cal H} = 4 \frac{G_{l'l}}{\sqrt{2}}
\sum_{n,\epsilon,\omega}          
g^n_{\epsilon\omega}   
\left[ \overline{l'_\epsilon} 
\Gamma^n {(\nu_{l'})}_\sigma \right]\, 
\left[ \overline{({\nu_l})_\lambda} \Gamma_n 
	l_\omega \right]\ ,
\label{eq:hamiltonian}
\ee
contains ten complex coupling constants or, since a common phase is
arbitrary, nineteen independent real parameters
which could be different for each leptonic decay.
The subindices
$\epsilon , \omega , \sigma, \lambda$ label the chiralities (left--handed,
right--handed)  of the  corresponding  fermions, and $n$ the
type of interaction: 
scalar ($I$), vector ($\gamma^\mu$), tensor 
($\sigma^{\mu\nu}/\sqrt{2}$).
For given $n, \epsilon ,
\omega $, the neutrino chiralities $\sigma $ and $\lambda$
are uniquely determined.

Taking out a common factor $G_{l'l}$, which is determined by the total
decay rate, the coupling constants $g^n_{\epsilon\omega}$
are normalized to \cite{FGJ:86}
\bel{eq:normalization}
1 = \sum_{n,\epsilon,\omega}\, |g^n_{\epsilon\omega}/N^n|^2 \, ,
\ee
where 
$N^n 
=2$, 1,
$1/\protect\sqrt{3} $ for $n =$ S, V, T.
In the SM, $g^V_{LL}  = 1$  and all other
$g^n_{\epsilon\omega} = 0 $.

The couplings $g^n_{\epsilon\omega}$ can be investigated through the
measurement of the final charged--lepton distribution 
and with the inverse decay
$\nu_{l'} l\to l' \nu_l$. 
For $\mu$ decay, where precise measurements of the polarizations of
both $\mu$ and $e$ have been performed, 
there exist \cite{pdg:96}
stringent upper bounds on the couplings involving right--handed helicities.
These limits show nicely 
that the bulk of the $\mu$--decay transition amplitude is indeed of
the predicted V$-$A type:
$g^V_{LL}  > 0.96$ (90\% CL).
Improved measurements of the $\mu$--decay parameters will
be performed at PSI and TRIUMPH \cite{guill}.

\begin{figure}[hbt]
\centerline{\epsfxsize =7.0cm \epsfbox{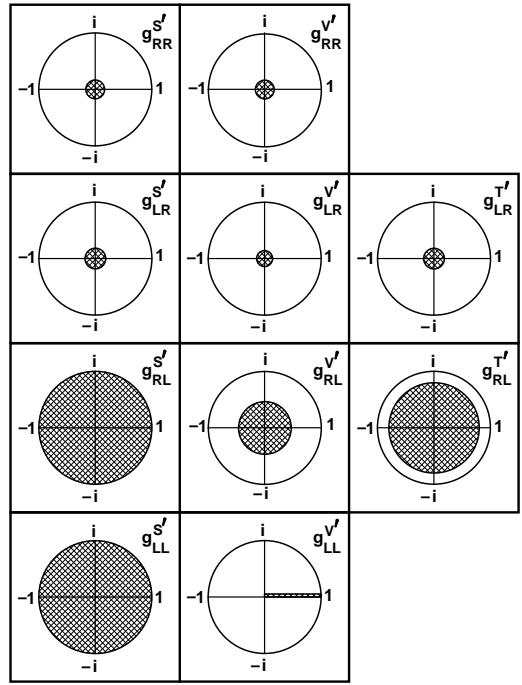}}
\vspace{-0.5cm}
\caption{90\% CL experimental limits \protect\cite{cleo:97a}
for the normalized $\tau$--decay couplings
$g'^n_{\epsilon\omega }\equiv g^n_{\epsilon\omega }/ N^n$,
assuming $e/\mu$ universality.}
\label{fig:tau_couplings}
\end{figure}

The $\tau$--decay experiments are starting to provide
useful information on the decay structure.
Figure \ref{fig:tau_couplings} shows the most recent limits
obtained by CLEO \cite{cleo:97a}.
The measurement of the $\tau$ polarization allows to bound those couplings
involving an initial right--handed lepton; however, information on the
final charged--lepton polarization is still lacking. Moreover,
the measurement of the inverse decay
$\nu_\tau l\to\tau\nu_l$, needed to separate the $g^S_{LL}$ and
$g^V_{LL}$ couplings, looks far out of reach.

\section{SEMILEPTONIC DECAYS}

From now on we will assume that the charged--current interaction
is indeed universal and of the V$-$A type, as predicted by the SM.

Let us consider the semileptonic weak decay
$H\to H' l^- \bar\nu_l$, associated with the corresponding quark transition
$d_j\to u_i l^-\bar\nu_l$.
The decay amplitude
\bel{eq:T_decay}
T  
\approx {G_F\over\sqrt{2}} \,\bV_{\!\! ij}\;
\cM^\mu_{H'H}\;
\bar l \gamma_\mu (1-\gamma_5) \nu_l
\ee
involves an hadronic matrix element of the quark current:
$\cM^\mu_{H'H}\equiv \langle H'| \bar u_i \gamma^\mu (1-\gamma_5) d_j |
H\rangle $. 
The evaluation of this matrix element is a non-perturbative QCD
problem and, therefore, introduces unavoidable theoretical uncertainties.

Usually, one looks for a semileptonic transition
where the matrix element can be fixed at some kinematical point, by
a symmetry principle 
This has the virtue of reducing the theoretical
uncertainties to the level of symmetry--breaking corrections and
kinematical extrapolations.
The standard example is a $0^-\to 0^-$ decay such as $K\to\pi l\nu$,
$D\to K l\nu$ or $B\to D l\nu$.
Only the vector current can contribute in this case:
\bel{eq:vector_me}
\cM^\mu_{P'P} \sim
(k+k')^\mu f_+(q^2) + (k-k')^\mu f_-(q^2) \, ,
\ee
where $q^2=(k-k')^2$ is
the momentum transfer carried by the intermediate $W$.
The unknown strong dynamics is fully contained in the two form factors
$f_\pm(q^2)$.

Since
$(k-k')^\mu\,\bar l\gamma_\mu (1-\gamma_5)\nu_l\sim m_l$, the contribution
of $f_-(q^2)$ is kinematically suppressed in the $e$ and $\mu$ modes.
Moreover, 
there is an additional mass suppression of the $f_-(q^2)$ term
for light quarks:
$f_-(q^2)\approx (m_{u_i}-m_{d_j})$.
The decay width can then be written as
\bel{eq:decay_width}
\Gamma  
 \approx {G_F^2 M_P^5\over 192\pi^3}\, |\bV_{\!\! ij}|^2\, 
|f_+(0)|^2\, \cI\, \left(1+\delta_{\mbox{\rms RC}}\right) ,
\ee
where $\delta_{\mbox{\rms RC}}$ is an electroweak radiative
correction factor and $\cI$ denotes a phase--space integral, 
which in the $m_l=0$ limit takes
the form
\bel{eq:ps_integral}
\cI\approx\int  
\, {dq^2\over M_P^8}\;
\lambda^{3/2}(q^2,M_P^2,M_{P'}^2)\, 
\left| {f_+(q^2)\over f_+(0)}\right|^2 .
\ee

The usual procedure to determine $|\bV_{\!\! ij}|$ involves three steps
\cite{PI:96b}:
\begin{enumerate}
\item Measure the shape of the $q^2$ distribution. This fixes the ratio
$|f_+(q^2)/f_+(0)|$ and therefore determines $\cI$.
\item Measure the total decay width $\Gamma$. Since $G_F$ is already known
from $\mu$ decay, the product $|f_+(0)|\, |\bV_{\!\! ij}|$ is determined.
\item Get a theoretical prediction for the normalization $f_+(0)$.
\end{enumerate}
The important point to realize is that theoretical input is
always needed. Thus, the accuracy of the $|\bV_{\!\! ij}|$
determination is limited by our ability to  calculate the relevant
hadronic input.

\begin{table*}[tbh]
\caption{Direct $\protect\bV_{\!\! ij}$ determinations.
\label{tab:V_CKM}}
\vspace{0.2cm}
\begin{tabular*}{\textwidth}{@{}l@{\extracolsep{\fill}}lll}
\hline
CKM entry & Value & Source & Symmetry
\\ \hline
$|\bV_{\!\! ud}|$ & $0.9740\pm 0.0005$ & Nuclear $\beta$ decay 
\cite{HA:97} & Isospin \   ($\chi$PT)
\\
& $0.979\phantom{0}\pm 0.002$ &
$n\to p e^-\bar\nu_e$ \cite{pdg:96,MA:91} & Isospin \   ($\chi$PT)
\\ \hline 
$|\bV_{\!\! us}|$ & $0.2196\pm 0.0023$ & $K_{e3}$ \cite{LR:84} &
SU(3) \ \    ($\chi$PT)
\\
& $0.222\phantom{0}\pm 0.003$ & Hyperon decays \cite{pdg:96} & 
SU(3) \ \   ($\chi$PT)
\\ \hline
$|\bV_{\!\! cd}|$ & $0.224\phantom{0}\pm 0.016$ & $\nu d \to c X$ 
\cite{pdg:96}  & ---
\\ \hline
$|\bV_{\!\! cs}|$ & $1.01\pm 0.18$ & $D \to \bar K e^+\nu_e$  
\cite{pdg:96} & ---
\\ \hline
$|\bV_{\!\! ub}|$ & $(3.3\pm 0.2{\,}^{+0.3}_{-0.4}\pm0.7)\times 10^{-3}$ &
 $B^0\to \pi^- l^+\nu_l, \rho^- l^+\nu_l$ \cite{CLEO:96} &   ---
\\ \hline
$|\bV_{\!\! ub}/\bV_{\!\! cb}|$ & $0.08\pm 0.02$ &
 $b\to u l^-\bar\nu_l$  \   (end-point) \cite{pdg:96} &   ---
\\ \hline
$|\bV_{\!\! cb}|$ & $0.038\pm 0.003$ & $B\to D^* l\bar\nu_l$ 
\cite{NE:97,berkelman,marinelli} &
$M_b\to\infty$ \   (HQET)
\\
& $0.040\pm 0.004$ & $b\to c l\bar\nu_l$  \   (inclusive)
\cite{NE:97,berkelman} &
$M_b\to\infty$ \   (HQET)
\\ \hline 
${|\bV_{\!\! tb}|\over
\sqrt{|\bV_{\!\! td}|^2 + |\bV_{\!\! ts}|^2 + |\bV_{\!\! tb}|^2}}$ \ \mbox{}
& $0.99\pm 0.29$ & $t\to b W / q W$ \cite{HE:97}& ---
\phantom{$\bigl\{ 
   \sqrt{{{\bV}_D}\over \sqrt{\overline{\bV}_D}}\bigr\}$}
\\  \hline
\end{tabular*}
\end{table*}

The present (direct) determinations of the CKM matrix elements are
summarized in Table~\ref{tab:V_CKM}.
For light quarks (u, d, s), the chiral symmetry of massless
QCD fixes the normalization of the relevant hadronic form factors at
zero momentum transfer; moreover, symmetry--breaking corrections
can be investigated rigorously with Chiral Perturbation
Theory ($\chi$PT) techniques \cite{PI:95}. 
Therefore, a rather good accuracy
has been achieved. Note however, that there is a long--standing
discrepancy ($\sim 2.5\,\sigma$) between the $|\bV_{\!\! ud}|$
values obtained from nuclear $\beta$ decay and the neutron lifetime
determination.

In the limit of infinite (c, b) quark masses the QCD Lagrangian has
additional flavour and spin symmetries \cite{IW:89}, which allow to fix the
normalization of the $\cM^\mu_{DB}$ and $\cM^\mu_{D^*B}$
hadronic matrix elements at the point of zero recoil (maximum momentum
transfer through the $W$ propagator). This point corresponds to the
kinematical configuration where the initial and final mesons have identical
velocities. Symmetry--breaking corrections can also be estimated with
the methods of Heavy Quark Effective Theory (HQET) \cite{NE:94}.
A reasonable determination of $|\bV_{\!\! cb}|$ can then be obtained.
Also shown in Table~\ref{tab:V_CKM} is the determination of this
CKM matrix element from the inclusive measurement of
$\Gamma(b\to c l\bar\nu_l)$; although free
from hadronic form factor uncertainties, this observable is very
sensitive to the not so well--known values of the bottom and charm
quark masses.

The remaining CKM determinations cannot make use of any useful symmetry
to control the hadronization effects (the relevant quarks are
too heavy to consider the $m_q\to 0$ limit, and/or too light for the
$m_q\to\infty$ approximation to make sense). Those determinations
need to rely on explicit hadronic models; thus, the achievable
precision is strongly limited by theoretical uncertainties.
An obvious exception is the recent constraint on
$|\bV_{\!\! tb}|$, obtained at the Tevatron.

\subsection{CKM Unitarity}

The values of $|\bV_{\!\! ui}|$ ($i=d,s,b$) provide
a test of the unitarity of the CKM matrix:
\bel{eq:unitarity_test}
|\bV_{\!\! ud}|^2 + |\bV_{\!\! us}|^2 + |\bV_{\!\! ub}|^2 \, = \,
0.9973\pm 0.0013 \, .
\ee
To get this number, we have used the weighted average of the two
$|\bV_{\!\! us}|$ determinations and the nuclear $\beta$ decay measurement
of $|\bV_{\!\! ud}|$.
Given the disagreement with the neutron lifetime determination,
it looks quite plausible that the 
small unitarity violation in Eq.~\eqn{eq:unitarity_test} originates
in the input $|\bV_{\!\! ud}|$ value.

Assuming unitarity, a more precise picture of the mixings among
the three quark generations is obtained \cite{pdg:96}.
The CKM matrix shows a hierarchical pattern, with the
diagonal elements being very close to one, the ones connecting the
two first generations having a size
\bel{eq:lambda}
\lambda\equiv |\bV_{\!\! us}| = 0.2205\pm 0.0018 \, ,
\ee
the mixing between the second and third families being of order
$\lambda^2$, and the mixing between the first and third quark flavours
having a much smaller size of about $\lambda^3$.
It is then quite practical to use the 
approximate parameterization \cite{WO:83}:
$$
\bV =\left[ \!\!
\begin{array}{ccc}
1- {\lambda^2 \over 2} & \lambda & A\lambda^3(\rho  - i\eta) \\
-\lambda & 1 -{\lambda^ 2 \over 2} & A\lambda^ 2 \\
A\lambda^ 3(1-\rho -i\eta) & -A\lambda^ 2 & 1 \end{array}
\!\!\right] \! ,
$$
valid up to $\cO(\lambda^ 4)$ corrections.
Here,
\beqn\label{eq:circle}
\lefteqn{A= {|\bV_{\!\! cb}|\over\lambda^2} = 0.80\pm 0.06 \, ,}\\
\lefteqn{\sqrt{\rho^2+\eta^2} \, = \, 
\left|{\bV_{\!\! ub}\over \lambda \bV_{\!\! cb}}\right|
\, =\, 0.36\pm 0.09 \, .}
\eeqn

\section{NON-LEPTONIC TRANSITIONS}

The dynamical effect of the strong interaction is more important in
non-leptonic transitions, where two different quark currents
are involved and gluons can couple everywhere.
Using the operator product expansion and renormalization--group
techniques, these transitions can be described through
effective Hamiltonians of the form
\bel{eq:eff_ham}
\cH_{\mbox{\rms eff}} = \sum_i C_i(\mu)\, Q_i \, ,
\ee
with local four--fermion operators $Q_i$, modulated
by Wilson coefficients $C_i(\mu)$.
The arbitrary renormalization scale $\mu$ separates the
short-- ($M >\mu$) and long-- ($M <\mu$) distance contributions,
which are contained in $C_i(\mu)$ and $\langle Q_i\rangle(\mu)$
respectively. 
Thus, $C_i(\mu)$ contain all the information on CKM factors and
heavy--mass scales.
The physical amplitudes are of course 
independent of $\mu$.

A lot of effort has been invested recently in the calculation
of Wilson coefficients at next-to-leading order. For the most
important processes, all contributions of
$\cO(\alpha_s^n L^n)$ and $\cO(\alpha_s^{(n+1)} L^n)$
have been computed, where $L=\log{(M/m)}$ denotes the logarithm
of any ratio of heavy--mass scales ($M,m\geq\mu$).
Moreover, the full $m_t/M_W$ dependence (at lowest order in $\alpha_s$)
has been also included.
A detailed summary of those calculations (with a complete
list of references) can be found in Ref.~\cite{BF:97}.

Unfortunately, in order to predict the physical amplitudes one is still
confronted with the calculation of the hadronic matrix elements 
$\langle Q_i\rangle(\mu)$
of the four--fermion operators $ Q_i$. This is a very difficult
non-perturbative problem which so far remains unsolved.
We have only been able to obtain rough estimates using
different approximations (vacuum saturation, $N_C\to\infty$ limit, 
QCD low--energy effective action, \ldots)
or applying QCD techniques (lattice, QCD sum rules) which suffer from
their own technical limitations.

\section{$B^0$--$\bar B^0$ MIXING}

Additional information on the CKM parameters
can be obtained from
flavour--changing neutral--current transitions, occurring at the 1--loop
level. An important example is provided by 
the mixing between the $B^0$ meson and its antiparticle.
This process occurs through the so-called box diagrams,
where  two $W$ bosons are  
exchanged between a pair of quark lines.
The mixing amplitude is proportional to
$$
\langle\bar B_d^0 | \cH_{\Delta B=2}|B_0\rangle\,\sim\,
\sum_{ij}\, \bV_{\!\! id}^{\phantom{*}}\bV_{\!\! ib}^*
\bV_{\!\! jd}^*\bV_{\!\! jb}^{\phantom{*}}\,
S(r_i,r_j) ,
$$
where $S(r_i,r_j)$ is a loop function which depends on the masses
[$r_i\equiv m_i^2/M_W^2$] 
of the up-type quarks running along the internal fermionic lines.

\begin{figure}[t]  
\centerline{\mbox{\epsfxsize=7.5cm\epsffile{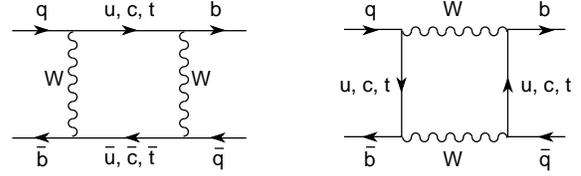}}}
\vspace{-0.5cm}
\caption{$B^0$--$\bar B^0$ mixing diagrams.}
\label{fig:boxdia}
\end{figure}

Owing to the unitarity of the CKM matrix, the mixing amplitude vanishes
for equal (up-type) quark masses (GIM mechanism \cite{GIM:70}); 
thus the effect
is proportional to the mass splittings between the $u$, $c$ and $t$ quarks.
Since the different CKM factors have all a similar size,
$\bV_{\!\! ud}^{\phantom{*}}\bV_{\!\! ub}^*\sim
\bV_{\!\! cd}^{\phantom{*}}\bV_{\!\! cb}^*\sim
\bV_{\!\! td}^{\phantom{*}}\bV_{\!\! tb}^*\sim A\lambda^3$, 
the final amplitude
is completely dominated by the top contribution;
i.e. $\langle\bar B_d^0 | \cH_{\Delta B=2}|B_0\rangle\,\sim\, 
|\bV_{\!\! td}|^2 S(r_t,r_t)$.
This transition can then be used to perform an indirect determination
of $|\bV_{\!\! td}|$.

The main theoretical uncertainty stems from the hadronic
matrix element of the $\Delta B=2$ four--quark operator
generated by the box diagrams:
$$     
\langle\bar B^0\, |\, (\bar b_L\gamma^\mu d_L)\: 
(\bar b_L\gamma_\mu d_L)\, |\, B^0\rangle \,\equiv\,
 {2\over 3} \, M_B^2\, (\sqrt{2}\, \xi_B)^2 \, .
$$     
The size of this matrix element is characterized through
the non--perturbative parameter
$\xi_B\equiv f_B\sqrt{\hat B_B}$, which is rather badly known.
Lattice calculations give \cite{FL:96}
$\sqrt{2} \xi_B = 207\pm 30$ MeV, while QCD sum rules provide
a slightly larger (but consistent) value,
$\sqrt{2} \xi_B = 260\pm 70$ MeV \cite{PP:95}.
Taking the range
$\sqrt{2} \xi_B = 215\pm 40$ MeV,
the measured mixing between the $B_d^0$--$\bar B_d^0$ mesons,
$\Delta M_{B^0_d} = 0.463\pm 0.018$ ps$^{-1}$  \cite{berkelman,barberio},
implies:
\bel{eq:V_td}
 |\bV_{\!\! td}|\, = \, 0.0080\pm 0.0003\, {}^{+0.0018}_{-0.0013} \,
, 
\ee
where the first error is the experimental one and the second reflects
the theoretical uncertainties.

In terms of the $(\rho,\eta)$ parameterization,
this gives
\bel{eq:circle_t}
\sqrt{(1-\rho)^2+\eta^2} \, = \,
\left|{\bV_{\!\! td}\over \lambda\bV_{\!\! cb}}\right|
\, = \, 0.93\, {}^{+0.22}_{-0.17} \, .
\ee

A similar analysis can be applied to the $B^0_s$--$\bar B^0_s$ mixing
probability. The non--perturbative uncertainties can be reduced to the
level of $SU(3)$ breaking corrections through the ratio
\bel{x_ratio}
{\Delta M_{B^0_s}\over \Delta M_{B^0_d}} \approx
{M_{B^0_s}\, \xi^2_{B^0_s}\over M_{B^0_d}\, \xi^2_{B^0_d}}\,
\left|{\bV_{\!\! ts}\over \bV_{\!\! td}}\right|^2
\equiv \Omega \times 
\left|{\bV_{\!\! ts}\over \bV_{\!\! td}}\right|^2 \, .
\ee
Taking
$\Omega\approx 1.15\pm 0.15$,
the present bound $\Delta M_{B^0_s} > 10.2$ ps$^{-1}$ 
(95\%\,\mbox{\rm CL}) \cite{berkelman,crawford}
implies
\bel{eq:Vts_Vtd}
\left|{\bV_{\!\! ts}\over \bV_{\!\! td}}\right|
\approx {1\over \lambda  \sqrt{(1-\rho)^2+\eta^2}} > 3.8 \; .
\ee

\section{CP VIOLATION}

With only two fermion generations the quark--mixing mechanism cannot
give rise to CP violation; therefore,
for CP violation to occur in a particular process,
all 3 generations are required to play an active role.
In the kaon system, for instance, CP--violation effects can only
appear at the one--loop level, where the top quark is present.
In addition, all CKM matrix elements must be non--zero and the quarks
of a given charge must be non--degenerate in mass. If any of these
conditions were not satisfied, the CKM phase could be rotated away
by a redefinition of the quark fields. CP--violation effects
are then necessarily proportional to the product of all CKM angles, and
should vanish in the limit where any two (equal--charge) quark masses
are taken to be equal.
Thus, violations of the CP symmetry are necessarily small.

Up to know, the only experimental evidence of CP--violation phenomena
comes from the kaon system. 
The ratios,
\beqn\label{eq:etapm}
\eta_{+-} &\!\! \equiv &\!\! {A(K_L\to\pi^+\pi^-)\over A(K_S\to\pi^+\pi^-)}
  \,\approx\, \varepsilon_K^{\phantom{'}}
 + \varepsilon_K'          
  \, , \\ \label{eq:etazero}
\eta_{00} &\!\! \equiv &\!\!  {A(K_L\to\pi^0\pi^0)\over A(K_S\to\pi^0\pi^0)}
  \,\approx\, \varepsilon_K^{\phantom{'}} 
- 2\varepsilon_K'         
\, ,\quad
\eeqn
involve final $2\pi$  states which are even under CP. Therefore, they
measure a CP--violating amplitude which can originate
either from a small CP--even admixture in the initial $K_L$ state
(indirect CP violation), parameterized by $\varepsilon_K$, or from
direct CP violation in the decay amplitude. This latter effect,
parameterized by $\varepsilon_K'$, requires the interference between
the two $K\to 2\pi$ isospin ($I=0,2$)  amplitudes, with
different weak and strong phases.

The parameter $\varepsilon_K$ is well determined \cite{pdg:96}:
\beqn\label{eq:eps_K}
\lefteqn{\varepsilon_K  =   (2.280\pm 0.013)\times 10^{-3}\; 
     \e^{i\phi(\varepsilon_K)} \, ,}
\\
\lefteqn{\phi(\varepsilon_K) \approx 
   \arctan{\left( {-2\Delta M_{K^0}\over\Delta\Gamma_{K^0}}\right)}
   = 43.49^\circ\pm0.08^\circ  .} \no
\eeqn
$\varepsilon_K$ has been also measured \cite{pdg:96} through the
CP asymmetry between the two
$K_L\to\pi^\mp l^\pm\bnul$ semileptonic decay widths, which implies 
Re$(\varepsilon_K) = (1.63\pm0.06)\times 10^{-3}$, in good agreement
with \eqn{eq:eps_K}.

The value of $\varepsilon_K'$ is not so-well established.
Two different experiments
have recently reported a measurement of this
quantity:
$$
\mbox{\rm Re}\left({\varepsilon'_K\over\varepsilon^{\phantom{'}}_K}\right) 
 = 
\left\{ \!
\begin{array}{ll} 
(23.0\pm 6.5)\times 10^{-4}  & \quad 
   [\mbox{\rm NA31}]  
  \\
(7.4\pm5.9)\times 10^{-4} & \quad 
   [\mbox{\rm E731}]  
\ea \right. .
$$
The NA31 measurement \cite{NA31:93} provides
evidence for a non--zero value of 
$\varepsilon'_K/\varepsilon^{\phantom{\prime}}_K$ (i.e.,
direct CP violation), with a statistical significance of more than three
standard deviations. However, this is not supported by
the E731 result \cite{E731:93}, which is compatible with 
$\varepsilon'_K/\varepsilon^{\phantom{\prime}}_K = 0$, 
thus with no direct CP violation.
The probability for the two results being statistically compatible is
only 7.6\%.

The next generation of 
$\varepsilon'_K/\varepsilon^{\phantom{\prime}}_K$ experiments is already
ready at CERN (NA48 \cite{NA48}) and Fermilab (KTEV \cite{KTEV}).
Moreover, a dedicated $\phi$ factory (DA$\Phi$NE), providing
large amounts of tagged $K_S$, $K_L$ and $K^\pm$
($\phi\to K\bar K$), will start running soon 
at Frascati \cite{DAPHNE}. The goal of all these experiments is to
reach sensitivities better than $10^{-4}$.

The CKM mechanism generates CP--violation effects both in the
$\Delta S=2\; $ $K^0$--$\bar K^0$ transition (box diagrams) and in the
$\Delta S=1$ decay amplitudes (penguin diagrams).
The theoretical analysis of $K^0$--$\bar K^0$ mixing is
quite similar to the one applied to the $B$ system. This time, however,
the charm loop contributions are non--negligible. The main
uncertainty stems from the calculation of the hadronic matrix
element of the four--quark $\Delta S=2$ operator, which is usually
parameterized through the non--perturbative parameter \cite{PP:95}
$\hat B_K\approx 0.4$--0.8.

The experimental value of $\varepsilon_K$ specifies a hyperbola in the
$(\rho,\eta)$ plane. 
This is shown in Fig.~\ref{fig:unitarity_constraints},
together with the constraints obtained from
$|\bV_{\!\! ub}/\bV_{\!\! cb}|$ and $B_d^0$--$\bar B_d^0$ mixing.
This figure, taken from Ref.~\cite{BF:97}, assumes
$\hat B_K = 0.75\pm 0.15$ and
$\sqrt{2}\xi_B = 200\pm 40$ MeV.
Also shown in the figure is the impact of the experimental bound on
$\Delta M_{B_s^0}$.  

%
\begin{figure}[t]  
\centerline{\rotate[r]{\epsfysize =7.5cm \epsfbox{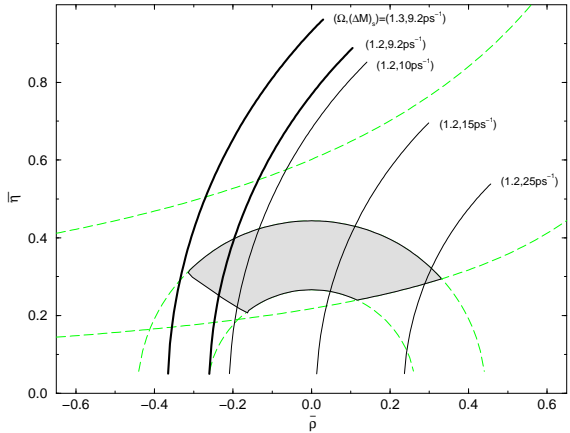}}}
\vspace{-0.5cm}
\caption{Unitarity Triangle constraints
\protect\cite{BF:97}.}
\label{fig:unitarity_constraints}
\end{figure}
%

The theoretical estimate of $\varepsilon'_K/\varepsilon^{\phantom{\prime}}_K$
is much more involved, because ten  
four--quark operators need to be considered
in the analysis and the presence of cancellations between different
contributions tends to amplify the sensitivity to the not
very well controlled long--distance effects.
For large values of the top mass, the $Z^0$--penguin contributions
strongly suppress the expected value of 
$\varepsilon'_K/\varepsilon^{\phantom{\prime}}_K$,
making the final result very sensitive to $m_t$.
The present theoretical estimates \cite{BF:97} 
range from $-1.2\times 10^{-4}$ to $1.6\times 10^{-3}$.
More theoretical work is needed in order to get firm predictions.

\subsection{The Unitarity Triangle}

The SM mechanism of CP violation is based in the unitarity of the
CKM matrix. 
Up to now, the only unitarity relation which has been precisely tested
is the one associated with the first row;  
however, only the moduli of the CKM parameters appear in
Eq.~\eqn{eq:unitarity_test}, while CP violation has to do with their phases.
The most interesting off--diagonal unitarity condition is
\bel{eq:triangle}
\bV^\ast_{\!\! ub}\bV^{\phantom{*}}_{\!\! ud}  + 
\bV^\ast_{\!\! cb}\bV^{\phantom{*}}_{\!\! cd}  + 
\bV^\ast_{\!\! tb}\bV^{\phantom{*}}_{\!\! td}  =  0 \, ,
\ee
which involves three terms of similar size.
This relation can be visualized by a triangle in the complex
plane, which is usually
scaled by dividing its sides by 
$\bV^\ast_{\!\! cb}\bV^{\phantom{*}}_{\!\! cd}$.
This aligns one
side of the triangle along the real axis and makes its length equal to
1; the coordinates of the 3 vertices are then
$(0,0)$, $(1,0)$ and $(\rho,\eta)$.
In the absence of CP violation, this unitarity triangle would degenerate
into a segment along the real axis.

Although the orientation of the triangle in the complex plane
is phase--convention dependent, the triangle itself is a physical
object: the length of the sides and the angles 
($\alpha$, $\beta$, $\gamma$) can be directly
measured.
In fact, we have already determined its sides from 
$\Gamma(b\to u)/\Gamma(b\to c)$ and   
$B^0_d$--$\bar B^0_d$ mixing, and the position of the $(\rho,\eta)$
vertex has been further pinned down with the additional information
provided by $\varepsilon_K$.

\section{FUTURE PROSPECTS}

\subsection{Bottom Decays}


The flavour--specific decays
$B^0\to X l^+\nu_l$ and $\bar B^0\to X l^-\bar\nu_l$
provide the most direct way to measure the amount of CP violation in
the $B^0$--$\bar B^0$ mixing matrix, through
the asymmetry between the number of $l^+l^+$ and $l^-l^-$ pairs produced
in the processes $e^+e^-\to B^0\bar B^0\to l^\pm l^\pm X$.
This $\Delta B = 2$ asymmetry is expected to be quite tiny
in the SM, because
$|\Delta\Gamma_{B^0}/\Delta M_{B^0}|
\sim m_b^2/m_t^2 << 1 \,\,$;
 moreover, there is an additional GIM suppression
$\sim (m_c^2-m_u^2)/ m_b^2$.
The observation of an asymmetry  at the percent level,
would be a clear indication of new physics beyond the SM.

Direct CP violation could be established by measuring a non-zero
rate asymmetry in $B^\pm$ decays. Unfortunately, the necessary
presence of a strong--phase difference makes difficult to obtain
clean information on the CKM matrix from this kind of observables.

The large $B^0$--$\bar B^0$ mixing provides a different way to generate the
required CP--violating interference.
There are quite a few non--leptonic final states which are reachable
both from a $B^0$ and a $\bar B^0$. For these flavour non--specific decays
the $B^0$ (or $\bar B^0$) can decay directly to the given final state $f$,
or do it after the meson has been changed to its antiparticle via the
mixing process; i.e., there are two different amplitudes,
$A(B^0\to f)$ and $A(B^0\to\bar B^0\to f)$, corresponding to two possible
decay paths. CP--violating effects can then result from the interference
of these two contributions.

$B^0$ decays into CP self--conjugate final states
are particularly promising. In that case,
assuming that only one weak amplitude contributes to the $B^0\to f$ 
transition,
all dependence on strong interaction effects
disappears  from the CP--violating rate asymmetries.   
Therefore, they could provide a direct and clean measurement of the CKM
parameters.
The angles of the unitarity triangle could be directly determined through
the decay modes 
$ B^0_d\to J/\psi K_S$ ($\beta$), $ B^0_d\to\pi^+\pi^-$ 
($\beta + \gamma = \pi - \alpha$)
and $ B^0_s\to\rho^0 K_S$ ($\gamma$).

The crucial assumption is that only one weak amplitude contributes to
a given decay, which obviously is not the case; the (usually) dominant
W--exchange decay amplitude gets corrected by diagrams with different
CKM structure, such as the so-called {\it penguins}.
The gold--plated exception is $ B^0_d\to J/\psi K_S$, since
all decay amplitudes share the same dependence on CKM factors to an
excellent approximation; thus, it will provide a very clean measurement
of 
$\beta\equiv -\arg(
  \bV^{\phantom{*}}_{\!\! cd}\bV^*_{\!\! cb}/
   \bV^{\phantom{*}}_{\!\! td}\bV^*_{\!\! tb})$.
   
In the case of $ B^0_d\to\pi^+\pi^-$,
penguin diagrams generate indeed a different CKM dependence,
but they are numerically suppressed allowing for an
approximate determination of
$\alpha\equiv -\arg(
   \bV^{\phantom{*}}_{\!\! td}\bV^*_{\!\! tb}/ 
   \bV^{\phantom{*}}_{\!\! ud}\bV^*_{\!\! ub})$.
The measurement of $\gamma$ with  $ B^0_s\to\rho^0 K_S$
is however not feasible;
the {\it direct} decay amplitude is colour suppressed, leading presumably to
a large (maybe dominant) penguin contamination.  

Many additional tests of the CKM matrix with $B$ decays have been
proposed. The rich variety of available decay modes provides ways to
circumvent the strong interaction complications through relations
(isospin, SU(3), $D^0$--$\bar D^0$ mixing) among different processes
or measuring the time evolution.
A detailed summary of recent work can be found in Ref.~\cite{BF:97}.

\subsection{Rare K Decays}

The decay $K^+\to\pi^+\nu\bar\nu$ is a well--known example of an allowed
process where long--distance effects play a negligible role. Thus,
this mode provides a good test of the CKM structure.
The decay process is dominated by short--distance loops ($Z$ penguin,
$W$ box) involving the heavy top quark, but receives also sizeable
contributions from internal charm--quark exchanges.
The resulting decay amplitude
is proportional to    
the hadronic matrix element of the $\Delta S=1$ vector
current, which (assuming isospin symmetry) can be obtained from
$K_{l3}$ decays.

The branching ratio is predicted to be in the range \cite{BF:97} 
Br$\sim (9.1\pm 3.2)\times 10^{-11}$,
to be compared with the recently reported   
 signal (1 event) \cite{E787:97}:
\bel{eq:pred_br}
\mbox{\rm Br}(K^+\to\pi^+\nu\bar\nu) =
(4.2\, {}^{+9.7}_{-3.5})\times 10^{-10} \, .
\ee

The CP--violating decay $K_L\to\pi^0\nu\bar\nu$ has been suggested
\cite{LI:89} as a good candidate to look for  pure
direct CP--violating transitions. 
The contribution coming from indirect
CP violation via $K^0$--$\bar K^0$ mixing is very small \cite{LI:89}:
Br$|_\varepsilon \sim 5 \times 10^{-15}$.
The decay proceeds almost entirely through direct CP violation
(via interference with mixing), and
is completely dominated by short--distance loop diagrams with top--quark
exchanges \cite{BF:97}:
\bel{eq:br_pnn0}
\mbox{\rm Br}(K_L\to\pi^0\nu\bar\nu) \approx 8.07\times 10^{-11}\,
A^4 \, \eta^2 \, r_t^{1.15} \, .
\ee
The present experimental upper bound \cite{WE:94},
$\mbox{\rm Br}(K_L\to\pi^0\nu\bar\nu) < 5.8\times 10^{-5}$
(90\% CL),
is still far away from the expected range \cite{BF:97}
\bel{eq:pred_br0}
\mbox{\rm Br}(K_L\to\pi^0\nu\bar\nu) =
(2.8\pm 1.7)\times 10^{-11} \, .
\ee
Nevertheless, the experimental prospects to reach the required sensitivity
in the near future look rather promising \cite{adler}.
The clean observation of just a single
unambiguous event would indicate the existence of CP--violating
$\Delta S = 1$ transitions.

Another promising mode is $K_L\to\pi^0 e^+e^-$. Owing to the
electromagnetic suppression of the 2$\gamma$ CP--conserving
contribution, this decay seems to be dominated by the CP--violating
one--photon emission amplitude. Moreover, the direct CP--violating
contribution is expected to be larger than the indirect one
\cite{PI:96}.

A recent overview of many other interesting rare K decays can be
found in Ref.~\cite{PI:96}. 

\section{SUMMARY}

The flavour structure of the SM is one of the main pending questions
in our understanding of weak interactions.
Although we do not know the reason of the observed family replication,
we have learnt experimentally that the number of SM generations is
just three (and no more). Therefore, we must study as precisely
as possible the few existing flavours, to get some hints on the
dynamics responsible for their observed structure.

The SM incorporates a mechanism to generate CP violation, through the
single phase naturally occurring in the CKM matrix.
This mechanism, deeply rooted into the unitarity structure of $\bV$,
implies very specific requirements for CP violation
to show up, which should be tested in appropriate
experiments.
The tiny CP asymmetry observed in the K system,
can be parameterized through the CKM phase; however, we do not have yet
an experimental verification of the CKM mechanism.
A fundamental explanation of the origin of this phenomena is
also lacking.

New and powerful flavour factories will become
operational very soon.
Many   interesting CP--violation signals are expected to be seen in the
near future.
 Large surprises may well be
discovered, probably giving the first hints of new physics and
offering clues to the problems of fermion--mass generation, quark
mixing and family replication.


\end{document}